\documentclass{paper}

\begin{document}

\title{The Evolution of Magnetic Fields in Galaxy Clusters}
\author{Matthias Bartelmann$^1$, Klaus Dolag$^1$, Harald Lesch$^2$\\
  $^1$ Max-Planck-Institut f\"ur Astrophysik, P.O.~Box 1317, D--85741
  Garching, Germany\\
  $^2$ Universit\"atssternwarte M\"unchen, Scheinerstr.~1, D--81679
  M\"unchen, Germany}
\date{\em invited talk at IAU~208, Tokyo, 2001}

\begin{abstract}

Cosmological simulations of magnetic fields in galaxy clusters show
that remarkable agreement between simulations and observations of
Faraday rotation and radio haloes can be achieved assuming that seed
fields of $\sim10^{-9}\,\mbox{G}$ were present at redshifts
$\sim15$--$20$. The structure of the seed field is irrelevant for the
final intracluster field. On average, the field grows exponentially
with decreasing cluster redshift, but merger events cause steep
transient increases in the field strength. Typical field-reversal
scales are of order $50\,h^{-1}\mbox{kpc}$. In most cases, the
intracluster fields are dynamically unimportant. Assuming secondary
electron models, the average structure of cluster radio haloes can
naturally be reproduced.

\end{abstract}

\maketitle

\section{Introduction}

\begin{figure}[ht]
\includegraphics[width=\hsize]{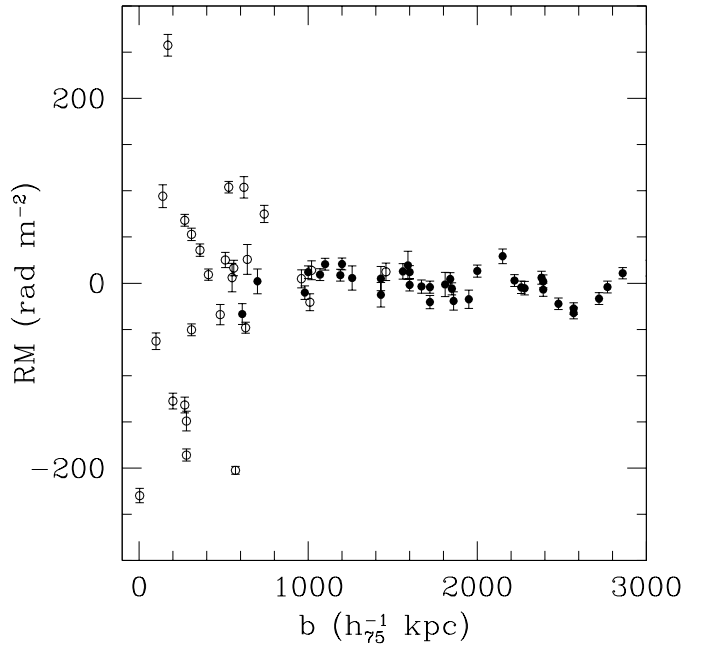}
\caption{Data by Clarke et al.~(2001) showing the Faraday rotation
measure in radio sources against their projected distance from the
nearest Abell cluster. While the {\em rms\/} rotation measure is
compatible with zero beyond $\sim1\,\mbox{Mpc}$, it increases
substantially towards smaller cluster-centric distances. This
demonstrates the existence of smooth intracluster magnetic fields.}
\label{fig:1}
\end{figure}

Magnetised plasmas are optically birefringent. The orientation of
polarised light traversing a magnetised plasma is therefore changed;
this effect is called Faraday rotation. Being proportional to the
square of the wavelength, it can be detected by observations of
polarised radio sources in two or more frequency bands. Numerous such
observations show that clusters contain magnetic fields which are
smooth, i.e.~ordered on scales comparable to the cluster scale
(e.g.~Vallee, MacLeod \& Broten 1986, 1987; Vallee 1990; Dreher et
al.~1987; Kronberg 1987; Kim et al.~1990; Kim, Kronberg \& Tribble
1991; Clarke, Kronberg \& B\"ohringer 2001 to name only a
few). Micro-Gauss field strengths are routinely
inferred. Figure~\ref{fig:1} illustrates data from Clarke et
al.~(2001). The Faraday rotation measure seen in polarised radio
sources is plotted against their projected separation from the nearest
Abell cluster. Evidently, the {\em rms\/} rotation measure increases
substantially towards cluster centres, while it is compatible with
zero at cluster-centric distances beyond $\sim1\,h^{-1}\mbox{Mpc}$.

This demonstrates that at least part of the Faraday rotation measured
in sources behind galaxy clusters is due to intracluster magnetic
fields rather than magnetic fields which are intrinsic to the sources
themselves.

The Faraday rotation measure is proportional to the integral along the
line-of-sight over the parallel component of the magnetic field, times
the electron density. The same rotation measure can therefore be
caused by strong fields which often change their direction, thus
causing substantial cancellation, or weak fields which are almost
homogeneous across a cluster. It is thus far from unique to infer
magnetic field strengths from Faraday-rotation measures. Typically,
fields with strengths of order $\mu\mbox{G}$ are inferred, assuming
simple magnetic-field geometries and typical electron densities
derived from cluster X-ray emission.

Another manifestation of intracluster magnetic fields is provided by
radio- and hard X-ray emission. Radio emission is caused by
relativistic electrons gyrating in the intracluster field. The same
population of relativistic electrons must Compton-upscatter photons of
the cosmic microwave background into the hard X-ray regime. From the
absence, or the low intensity, of this non-thermal, hard X-ray
emission, {\em lower\/} limits can be inferred on the intracluster
magnetic fields, because if the relativistic electron population is
poor, the magnetic field has to be strong in order to explain the
cluster radio haloes. These lower limits are typically also of
micro-Gauss order (e.g.~Bazzano 1990; Fusco-Femiano et al.~1999;
Bagchi et al.~1998; Henriksen 1999).

This outlines the motivation for our numerical work summarised
below. We wish to understand how magnetic fields evolve in galaxy
clusters, how they act back on their host clusters themselves, how the
initial conditions need to be set up so that observables like rotation
measures and radio haloes can be reproduced, and how the magnetic
fields are structured inside clusters.

\section{Numerical Technique}

For these simulations, we use the {\em GrapeMSPH\/} code (Dolag,
Bartelmann \& Lesch 1999), developed from M.~Steinmetz' (1996) {\em
GrapeSPH\/} code. The code uses the {\em Grape\/} hardware (Ito et
al.~1993) to trace dark-matter particles according to the softened
force law
\begin{equation}
  \frac{\mbox{d}\vec v_a}{\mbox{d}t}=-\sum_i
  \frac{m_i\,(\vec r_a-\vec r_i)}
       {(|\vec r_a-\vec r_i|^2+\epsilon_a^2)^{3/2}}\;,
\label{eq:1}
\end{equation}
where $a$ is the index of the particle under consideration, $\vec r_a$
and $\vec v_a$ are its position and velocity, and $\epsilon_a$ is its
softening length. The summation extends over all other particles at
positions $\vec r_i$ with masses $m_i$.

The dynamics of the gas is treated in the SPH approximation (Lucy
1977; Monaghan 1992). Particles are replaced by extended, soft spheres
whose density is described by a kernel function $W$ with variable
width $h$. Any physical quantity $A$ at position $\vec r$ is
approximated as
\begin{equation}
  \langle A(\vec r)\rangle\approx\sum_i\,
  m_i\,\frac{A(\vec r_i)}{\rho(\vec r_i)}\,
  W(\vec r-\vec r_i,h)\;,
\label{eq:2}
\end{equation}
where the sum extends over all particles which come sufficiently close
to $\vec r$. The code combines the advantages of SPH, which is a fast
method with adaptive resolution, with those of the {\em Grape\/}
hardware, which provides supercomputer performance on low-end
hardware. In our case, the host workstation is a two-processor SUN
Ultra-Sparc 2300 equipped with five {\em Grape\/}-3Af boards.

The equation of motion for the gas particles can symbolically be
written
\begin{eqnarray}
  \frac{\mbox{d}\vec v_a}{\mbox{d}t}&=&\mbox{pressure gradient}+
  \mbox{viscous stresses}+\mbox{magnetic force}\nonumber\\
  &+&\mbox{gravity}+\Omega_\Lambda\;.
\label{eq:3}
\end{eqnarray}
In particular, the back-reaction of the magnetic field on the gas flow
through the Lorentz force is included in the calculation. The term
$\Omega_\Lambda$ symbolises the additional acceleration in presence of
the cosmological constant. In addition to the momentum conservation
expressed by Eq.~(\ref{eq:3}), the energy equation is solved. The gas
is treated as an ideal gas with an adiabatic index of $5/3$. Cooling
and heating are foreseen in the code, but have so far been used for
testing purposes only.

We assume that the conductivity of the intracluster plasma is
infinite, so that the induction equation of ideal magnetohydrodynamics
can be used,
\begin{equation}
  \frac{\partial\vec B}{\partial t}=
  \nabla\times(\vec v\times\vec B)\;.
\label{eq:4}
\end{equation}
Equation~(\ref{eq:4}) and the inclusion of the Lorentz force in
Eq.~(\ref{eq:3}) constitute our main complements to the original
{\em GrapeSPH\/} code.

The numerical viscosity is chosen such as to minimise viscous angular
momentum transport in particle encounters (Monaghan \& Gingold 1983;
Balsara 1995; Steinmetz 1996). The viscosity vanishes exactly in pure
shear flows. Finally, all SPH expressions are symmetrised against
particle exchange.

\section{Simulations}

We tested the code extensively, as described in detail in Dolag et
al.~(1999). Specifically, the code succeeds in computing spherical
collapse models, and in solving shock tube problems (Brio \& Wu
1988). Although the code does not by construction conserve
$\nabla\cdot\vec B$, it turns out in our simulations that
$\nabla\cdot\vec B$ is always many orders of magnitude smaller than
$|\vec B|$ divided by the correlation length of $\vec B$.

We perform simulations in two different cosmological models, SCDM with
$\Omega_0=1$ and $\Omega_\Lambda=0$, and $\Lambda$CDM with
$\Omega_0=0.3$ and $\Omega_\Lambda=0.7$. In both cases, we initialise
dark matter distributions from the COBE-normalised CDM power
spectrum. Galaxy clusters are surrounded a layer of more massive
particles in order to mimic the gravitational tidal field exerted by
neighbouring large-scale structure. The total particle number is
limited by the {\em Grape\/} architecture, but our simulated clusters
are generally resolved by $\ge10^4$ particles.

\begin{figure}[ht]
\includegraphics[width=\hsize]{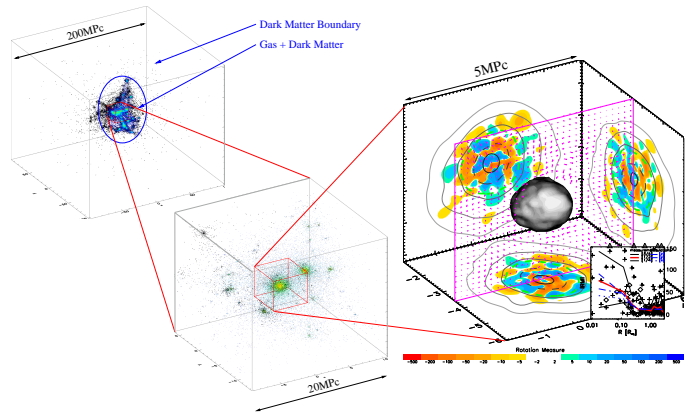}
\caption{One evolutionary stage of one of our cluster
simulations. Three boxes are shown which zoom into the cluster region
from left to right. The right-most box has a side length of
$5\,h^{-1}\mbox{Mpc}$. It shows a grey iso-density surface of the
intracluster gas distribution in the centre; the grey contours on the
box sides are isodensity curves of the intracluster gas. The coloured
patches on the box sides show the Faraday rotation measure, with
values ranging within $\pm500\,\mbox{rad}\,\mbox{m}^{-2}$. Red patches
have negative, blue patches positive rotation measure. The purple
frame in the middle of the box shows a slice through the intracluster
magnetic field, with the arrows indicating field strength and
orientation.}
\label{fig:2}
\end{figure}

Initial conditions are set up such that the average magnetic field
strength at high redshifts ($15$ to $20$ depending on cosmology) are
of nano-Gauss order. In lack of any detailed prediction of the
magnetic field structure at such redshifts, we use two vastly
different assumptions. The field is either assumed to be homogeneous
across the entire simulation volume, or chaotic, in which case its
amplitude is drawn from a pre-defined power spectrum, and its
direction is random, subject only to the condition that
$\nabla\cdot\vec B=0$. Figure~\ref{fig:2} shows one evolutionary stage
of one of our cluster simulations as an example.

The Faraday rotation measure created by the intracluster magnetic
field is show as the coloured patches projected onto the box
sides. The rotation measures range within
$\pm500\,\mbox{rad}\,\mbox{m}^{-2}$, increasing from red to
blue. Obviously, the rotation measures reach values typical for the
observations, they cover a substantial portion of the cluster's cross
section, and they show structure on scales of order
$\sim100\,h^{-1}\mbox{kpc}$. The figure also shows how the cluster is
embedded into the surrounding large-scale structure, and it displays
the strength and orientation of the magnetic field in a slice cut
through the cluster core.

\section{Results}

\subsection{Field Structure and Amplification}

\begin{figure*}[ht]
\includegraphics[width=0.48\hsize]{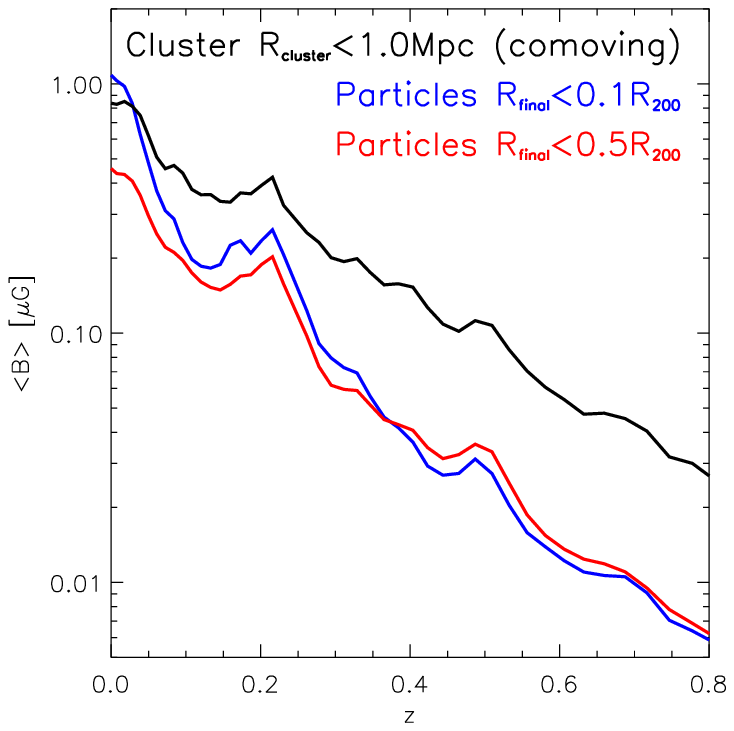}\hfill
\includegraphics[width=0.48\hsize]{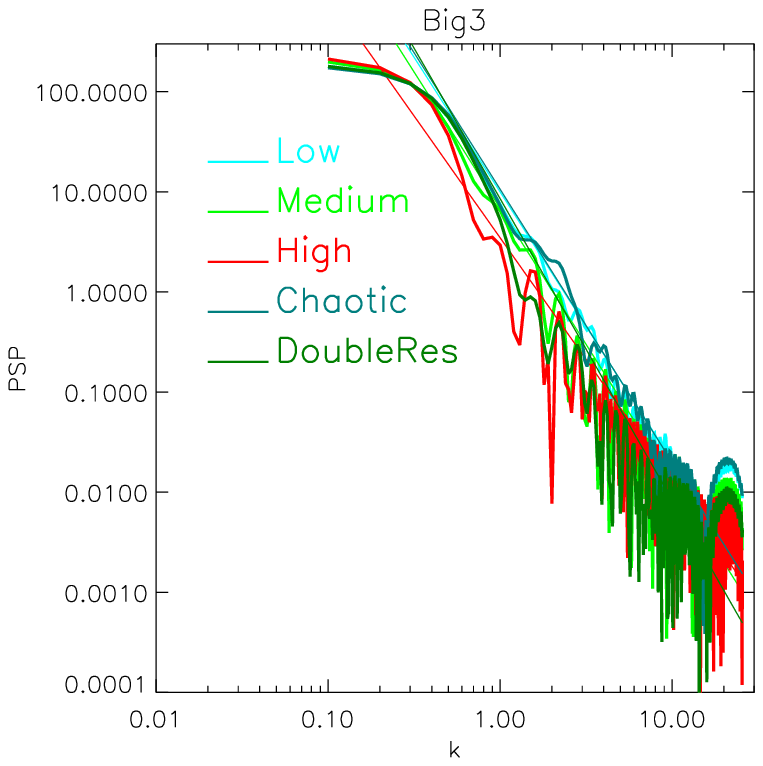}
\caption{{\em Left panel:\/} Growth with redshift of the magnetic
field strength averaged within many of our simulated clusters. On
average, the field strength increases exponentially with decreasing
redshift, approximately $\propto10^{-2z}$. -- {\em Right panel:\/}
Power spectra for the modulus of the intracluster magnetic field
strength for various initial field set-ups. Above the Nyquist
frequency of the simulation box, the power spectra are fairly steep
power laws with exponents between $-2.5$ and $-3$.}
\label{fig:3}
\end{figure*}

Starting from nano-Gauss fields at redshifts between $15$ and $20$,
intracluster fields reach micro-Gauss strength in and near cluster
centres at low redshift. Such field amplifications exceed expectations
from simple spherical collapse models of a gas cloud with frozen-in
magnetic field by about an order of magnitude. This becomes possible
due to shear flows in the intracluster gas, which stretch, bend and
entangle magnetic field lines and thereby increase the magnetic field
strength. The left panel in Fig.~\ref{fig:3} shows that the magnetic
field strength, averaged within the central regions of many clusters
in our simulated sample, increases approximately exponentially with
decreasing redshift, $\langle|\vec B|\rangle\propto10^{-2z}$ (Dolag et
al.~1999; Dolag, Bartelmann \& Lesch 2001).

We find that the initial field structure is entirely irrelevant for
the final field structure after cluster collapse. Simulations starting
from the homogeneous magnetic field set-up lead to final magnetic
fields which are statistically identical to those obtained from
chaotic initial fields. This shows that the final structure of
intracluster fields is completely determined by the cluster collapse,
in the course of which any information on the initial field structure
is erased.

The autocorrelation function of the magnetic field reveals typical
field-reversal scales of order $50\,h^{-1}\mbox{kpc}$. The right panel
in Fig.~\ref{fig:3} shows the power spectrum of the modulus of the
final magnetic field strength for various initial conditions. The
power spectrum resembles a power law for values of $k$ sufficiently
larger than the Nyquist frequency of the box, quite independent of the
initial field set-up. The power-law is fairly steep, with exponents
ranging within $-2.5$ to $-3$. This is steeper than expected for
Kolmogorov turbulence, but compatible with two-dimensional
magnetohydrodynamic turbulence. It cannot be excluded at this stage
that the numerical viscosity of the SPH code contributes to steepening
the magnetic-field power spectrum.

\subsection{Faraday Rotation}

Simulations in both cosmologies reproduce the statistics of observed
Faraday-rotation measures very well. On the whole, the amplitudes and
coherence lengths of the simulated rotation-measure maps, and their
projected radial profiles, agree excellently with the available
data. As noticed before, the initial field configuration is irrelevant
in that respect. Setting up fields with nano-Gauss strength at
redshifts $15$--$20$ is sufficient.

Merger events in the cluster history are prominently reflected in the
spatial distribution of the Faraday rotation measure. To illustrate
this point, we show in Fig.~\ref{fig:5} the fraction $\cal{F}$ of the
total cluster cross section which is covered by Faraday rotation
measures with absolute values exceeding given thresholds. The total
cluster cross section is defined as the area emitting 90\% of the
total cluster X-ray luminosity. In other words, we study the measure
of the excursion set of the Faraday rotation measure map.

\begin{figure}[ht]
\includegraphics[width=\hsize]{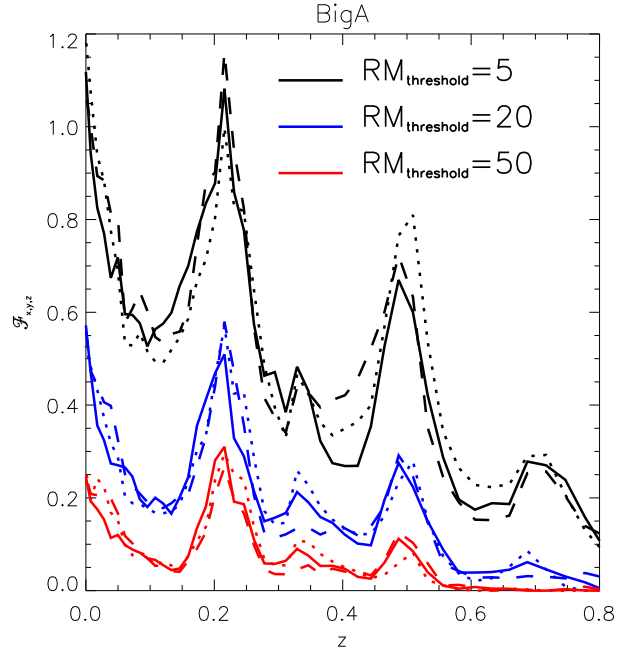}
\caption{Fraction $\cal{F}$ of the total cluster cross section covered
by Faraday rotation measures whose absolute values exceed the
thresholds indicated in the plot. The total cluster cross section is
defined to be the area emitting 90\% of the total cluster
luminosity. Irrespective of the threshold, the curves show steep,
transient increases of $\cal{F}$ on top of a more gentle increase.
The maxima coincide with the core passages following merger events.}
\label{fig:4}
\end{figure}

Figure~\ref{fig:4} shows a steady increase in $\cal{F}$ with
decreasing redshift, underlying steep, transient increases which are
independent of the threshold of the excursion set. Correlating these
increases with the cluster evolution shows that $\cal{F}$ increases
during merger events, reaching its maximum during core passage of the
merged substructure, and decreasing thereafter as the cluster
relaxes. Merger events therefore leave a pronounced imprint on the
appearance of a cluster in Faraday-rotation measurements.

Observationally, the {\em rms\/} rotation measure observed in galaxy
clusters is closely correlated with their X-ray surface
brightness. While this correlation contradicts simple analytic models
for the intracluster magnetic fields, it is naturally reproduced in
our simulated clusters (Dolag et al.~2001).

\subsection{Dynamics}

The intracluster magnetic fields provide pressure support adding to
the thermal gas pressure. Hydrostatic equilibrium with the cluster
potential wells can therefore be achieved with lower gas
temperatures. However, the fields in our simulated clusters are
typically so weak that the temperature reduction is of order a few per
cent only. Although it thereby increases the scatter in the
mass-temperature relation to some degree, this effect is generally
unimportant (Dolag, Evrard \& Bartelmann 2001).

\begin{figure}[ht]
\includegraphics[width=\hsize]{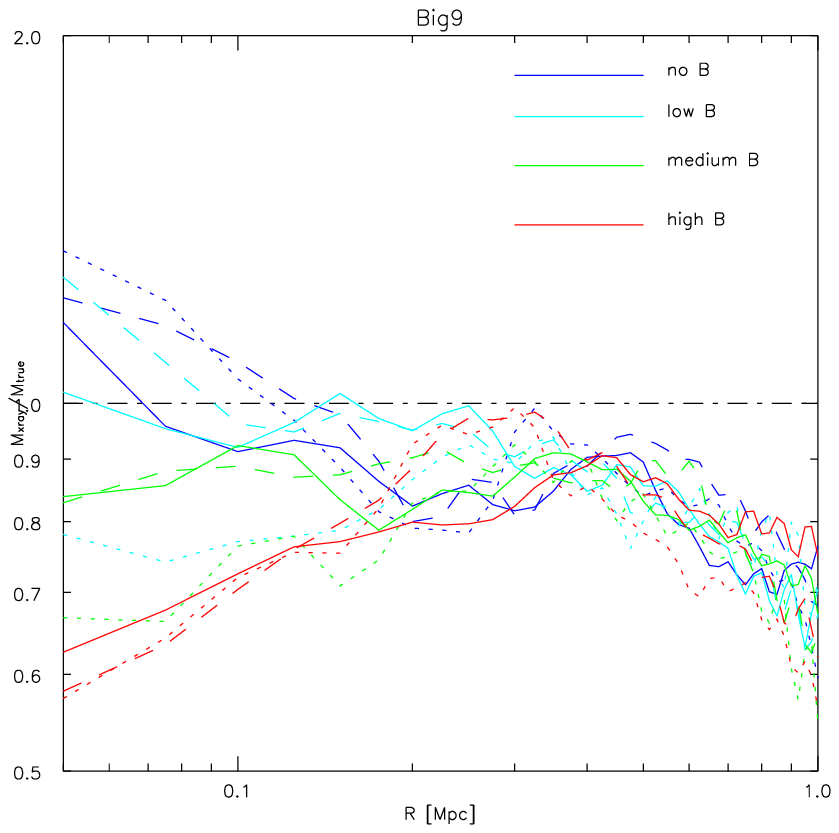}
\caption{Masses inferred from $\beta$ models fit to magnetised
clusters, divided by the true cluster masses, as a function of
cluster-centric radius. For clusters with comparatively strong fields,
the inferred mass near the cluster centres can drop to $\sim60\%$ of
its true value. Generally, however, the uncertainty is small compared
to the uncertainty in the $\beta$ model.}
\label{fig:5}
\end{figure}

Almost everywhere in the simulated clusters, and almost always during
their evolution, the magnetic fields are dynamically unimportant
compared to the uncertainties in the standard $\beta$ model commonly
used to infer cluster masses under the assumption of hydrostatic
equilibrium. Lower cluster temperatures cause $\beta$-model mass
estimates to be biased low. Figure~\ref{fig:5} shows that cluster
masses inferred from $\beta$ models to the X-ray emission of
magnetised galaxy clusters can be as low as $\sim60\%$ of the true
cluster masses, but only in cluster cores and if the fields are
comparatively strong. However, masses can substantially be
underestimated in highly substructured or merging clusters (Dolag \&
Schindler 2000).

\subsection{Radio Haloes}

\begin{figure}[ht]
\includegraphics[width=\hsize]{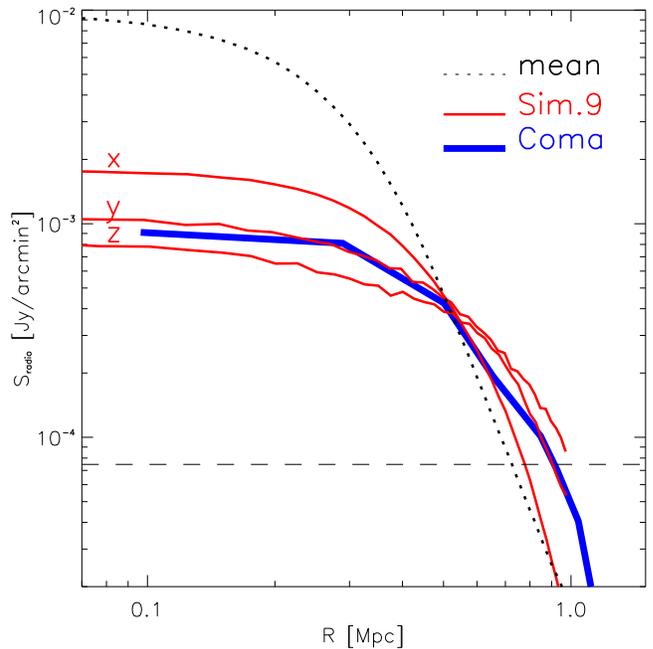}
\caption{Radial profiles of the radio surface brightness for the Coma
cluster (blue curve) and one of our simulated clusters assuming a
secondary electron model (red curves). While this particular
simulation reproduces Coma's radio-halo profile very well, the
simulated radio profiles are on average somewhat steeper.}
\label{fig:6}
\end{figure}

A natural consequence of intracluster magnetic fields is the
development of smooth radio emission on cluster scales, so-called
cluster radio haloes. Relativistic electrons in the intracluster
plasma gyrate in the magnetic field and emit synchrotron
radiation. Various models exist for the origin of the relativistic
electron population. They can be primarily accelerated, but it is more
likely that they from protons which are accelerated to relativistic
energies in intracluster shocks. These protons decay into relativistic
pions and electrons. Assuming such a secondary electron model, and
putting between 5\% and 15\% of the thermal energy content into
relativistic protons, our simulations can straightforwardly reproduce
the radio halo of the Coma cluster (see Fig.~\ref{fig:6}).

The figure shows the radial profile of the radio surface brightness in
one of our simulated clusters (red curves) and in the Coma cluster
(blue curve). While Coma's radio-halo profile is well reproduced by
one particular simulation, the mean profile averaged across our
complete simulated cluster sample is somewhat steeper. However, this
can easily be modified by assuming that different spatial
distributions of the relativistic protons from which the secondary
electrons originate.

Remarkably, though, the very steep correlation between the cluster
radio power and the X-ray temperature is naturally and accurately
reproduced by our simulations (Dolag \& En{\ss}lin 2000).

\section{Summary and Discussion}

We have performed cosmological simulations of magnetised galaxy
clusters in two CDM cosmogonies, SCDM and $\Lambda$CDM. The {\em
Grape\/} hardware allowed these simulations to be efficiently executed
on low-end hardware. We developed our {\em GrapeMSPH\/} code (Dolag et
al.~1999) by augmenting Steinmetz' (1996) {\em GrapeSPH\/} code with
the induction equation of ideal magnetohydrodynamics, and the Lorentz
force for the back-reaction of the field on the intracluster gas. The
code successfully passed an extensive suite of test runs.

Our cluster simulations lead to the following results:

\begin{itemize}

\item Micro-Gauss fields at low redshift are reached starting from
nano-Gauss fields at redshifts $15$--$20$, depending on the
cosmological model. Such field amplifications exceed by about an order
of magnitude expectations from magnetic flux conservation in the
simple spherical collapse model. They are possible due to shear flows
in the intracluster gas.

\item The initial structure of the magnetic fields is entirely
unimportant for the final structure. Simulations starting from
homogeneous magnetic fields lead to final fields which are
statistically identical to those obtained from chaotic initial
fields. The process of cluster formation wipes out any information on
the initial field configuration.

\item On average over many independent cluster simulations, the
magnetic field strength grows exponentially with decreasing redshift.

\item The simulations reproduce the observed statistics of
Faraday-rotation measures very well. Cluster merger events lead to
steep and transient increases in the area of the cluster which is
covered by substantial rotation measures. This area is maximised
during core passage of the merged substructure.

\item Almost everywhere in the simulated clusters, and almost always
during their evolution, magnetic fields are dynamically
unimportant. They cause some increased scatter in the relation between
X-ray temperature and mass. During merger events, however, cluster
masses inferred from $\beta$ fits can be lower than the true masses
by factors of two or more.

\item Assuming a secondary model for the production of relativistic
electrons, and putting $5\%$--$15\%$ of the thermal cluster energy
into relativistic protons, our simulated clusters have radio haloes
which agree well with observed haloes. In particular, the very steep
correlation between radio power and X-ray temperature is naturally
reproduced.

\end{itemize}

On the whole, it seems that our magnetohydrodynamic, cosmological
cluster simulations capture many features displayed by observed galaxy
clusters. However, the resolution of the simulations needs to be
increased before detailed and reliable predictions can be made on the
small-scale structure of intracluster magnetic fields and on the
structure and emissivity of radio haloes. At a later stage, cooling of
the intracluster gas has to be included, because it is expected to
change substantially the structure and dynamical importance of
magnetic fields in cluster cores.

\section*{Acknowledgments}

We thank the organisers of IAU~208 for the very interesting and
inspiring meeting. MB wishes to thank for generous financial support.

\end{document}